# Modular microfluidic platform for solubility measurement, nucleation statistics and polymorph screening of active pharmaceutical ingredients: Irbesartan, Rimonabant, Aripiprazole and Sulfathiazole

Mathilde Lambert [a, b], Romain Grossier [a]*, Mehdi Lagaize [a], Thirou Bactivelane [a], Vasile Heresanu[a], Benoît Robert [b], Nadine Candoni [a]*, Stéphane Veesler [a]*

[a] Centre Interdisciplinaire de Nanosciences de Marseille, CNRS, Aix-Marseille Université, CINaM-UMR7325, Campus de Luminy, Case 913, 13288 Marseille Cedex 09, France
[b] Sanofi R&D – Global CMC / Synthetics – Early Development France, Impasse des ateliers, 94400 Vitry-Sur-Seine, France
* Corresponding authors:
Tel.: +336 6292 2866; e-mail address: stephane.veesler@cnrs.fr (S. Veesler).
Tel.: +336 1724 8087; e-mail address: nadine.candoni@univ-amu.fr (N. Candoni),
Tel.: +336 6292 2879; e-mail address: romain.grossier@cnrs.fr (R. Grossier).

**Abstract**

Drug efficacy strongly relies on the solid state of the active pharmaceutical ingredient. Classical solid-state screening methods involve different solvent compositions and supersaturations. Moreover, the many repeat experiments needed to address the stochasticity of nucleation make this approach costly. This paper presents a newly developed modular microfluidic platform that provides a universal and flexible plug-and-play tool for crystallisation studies without use of surfactants. By dissolving a powder, our set-up generates saturated solutions that can be used for solubility measurements or distributed in microdroplets. Here, we describe solubility measurements performed on different forms, stable and metastable, of pharmaceutical molecules (Irbesartan, Rimonabant and Aripiprazole) in organic and aqueous solvents. In addition, we provide nucleation statistics obtained for Sulfathiazole in water and in acetonitrile. Reporting polymorph screening on Sulfathiazole and statistics for nucleated forms, we find that the cooling rate influences both nucleation and polymorphism results, reflecting the competition between thermodynamics and kinetics. Three unknown forms were discovered, with XRD patterns and Raman spectra that do not match any referenced form. We also demonstrate the limitations of microfluidics for crystallisation by cooling: reducing the crystalliser volume considerably increases nucleation induction time.

1. Introduction

Drug efficacy strongly relies on the solid state of the active pharmaceutical ingredient. Most of these organic molecules can exist under several crystalline structures (i.e. phases): polymorphs, solvates, salts, etc. [1], [2]. Here, the different crystalline structures will be termed "forms" (as common in the literature), without distinction. These forms have different physicochemical properties, which can affect the manufacturing process, the bioavailability and the posology of the drug [3], [4]. The crystallisation conditions of all forms and their relative stabilities must be known to avoid form transitions or appearance of new polymorphs, as in the case of Ritonavir in 1997 [5]. In the pharmaceutical industry, polymorphism studies really started after this event. Classical polymorph screening methods involve different solvent compositions and different supersaturations, using for example various cooling rates or non-solvents. All these screening experiments are usually carried out in crystallisers ranging from millilitres to hundreds of microlitres in capacity. Furthermore, due to the stochasticity of nucleation, each experiment (experimental condition) must be repeated a significant number of times [6]. Thus, each experiment requires 1 to 10 mg of raw material [7], making this approach costly.

One increasingly used solution to this problem is droplet-based microfluidics, involving a wide range of techniques [8]–[10]. Small-volume droplets, from nanolitre to microlitre, can be used as a single crystalliser, leading most of the time to a single nucleation event due to the confinement effect [11], [12]. Droplet lab-on-a-chip experiments are widely used for all sorts of applications in solution crystallisation [13]–[16]. However, there are some drawbacks to this approach. Once designed, microfluidic set-ups on chips are not adjustable. Moreover, a surface treatment needs to be applied when certain types of solvent are involved (aqueous, organic, etc). In addition, most applications require surfactants to generate droplets and prevent coalescence.

This paper presents the modular microfluidic platform we developed to provide a universal and flexible plug-and-play tool for crystallisation studies without use of surfactants. Improving the module assembly enables us to study a molecule of interest from the powder to the crystal, using image acquisition and spectroscopic characterisation. Our platform yields solubility measurements [17], monitors nucleation in droplets of 0.5μL (1mm diameter) using sequential image acquisition, and characterises crystal forms in-situ by Raman spectroscopy. This provides the percentage of nucleated crystals, as well as statistics on crystalline forms, thereby addressing the stochasticity of nucleation. The optical monitoring in time is also used to verify if any solution-mediated phase transition occurred. We present how the platform is applied to solubility measurements on various stable and metastable forms of Rimonabant, Sulfathiazole, Aripiprazole and Irbesartan. Then we present polymorph screenings on Sulfathiazole in water and acetonitrile, using different temperature profiles. Finally, we illustrate the influence of volume on nucleation by screening experiments on Rimonabant in different solvents.

## 2. Material
### 2.1. Products

- Rimonabant is a selective $CB_1$ receptor blocker developed by Sanofi-Aventis used for the treatment of obesity. Although now removed from the market, it is still studied for its polymorphism. Rimonabant has two known polymorphs with similar stabilities[18], I and II, and numerous solvates, principally in alcohols. Rimonabant was provided by Sanofi (SR141716, Form I: batch FFT.REX1.150.0031 / Form II: batch CL11469).

- Sulfathiazole is an antimicrobian agent widely studied in the literature for its polymorphism [19], [20]. However, there is confusion over the naming of each forms. We use the same labels as Munroe et al. for polymorphs I to V [19]. Sulfathiazole was purchased from Fluka Analytical (batch MKBQ0002V) and is a mixture of forms II and IV according to X-Ray Diffraction analysis (XRD) (see SIF).

- Among the drugs with the largest number of identified forms [21], Aripiprazole is an antipsychotic used to treat schizophrenia and bipolar disorders. Aripiprazole was purchased from Thermoscientific (batch A0417885) and is a mixture of forms II and III according to XRD analysis.

- Irbesartan is an Angiotensin-II Receptor blocker used in the treatment of hypertension. This organic molecule presents prototropic tautomerism due to reversible proton transfer on the tetrazole ring. The two resulting tautomers crystallise separately into two different crystalline structures, form A (1-H tautomer) and B (2-H tautomer)[22]. This rare type of phase transition is known as desmotropy. Both forms A and B were provided by Sanofi.

- Solvents are water, ethanol and acetonitrile, of analytical grade.

- Fluorinated oil GPL is a chemically inert oil, which is non miscible with most of the solvents. It was chosen due to its high viscosity, which reduces the risk of droplet coalescence, avoiding any use of surfactant. Fluorinated oil (GPL107, Krytox™) was purchased from Chemours; its viscosity at different temperatures, density and interfacial energy with air, water, ethanol and acetonitrile are presented in 0.

### 2.2. Microfluidic platform

Our microfluidic platform (Figure 1) is composed of different modules supported by an aluminium structure (MiniTec, Thorlabs) designed in-house. The modules can be organised as required for different applications.

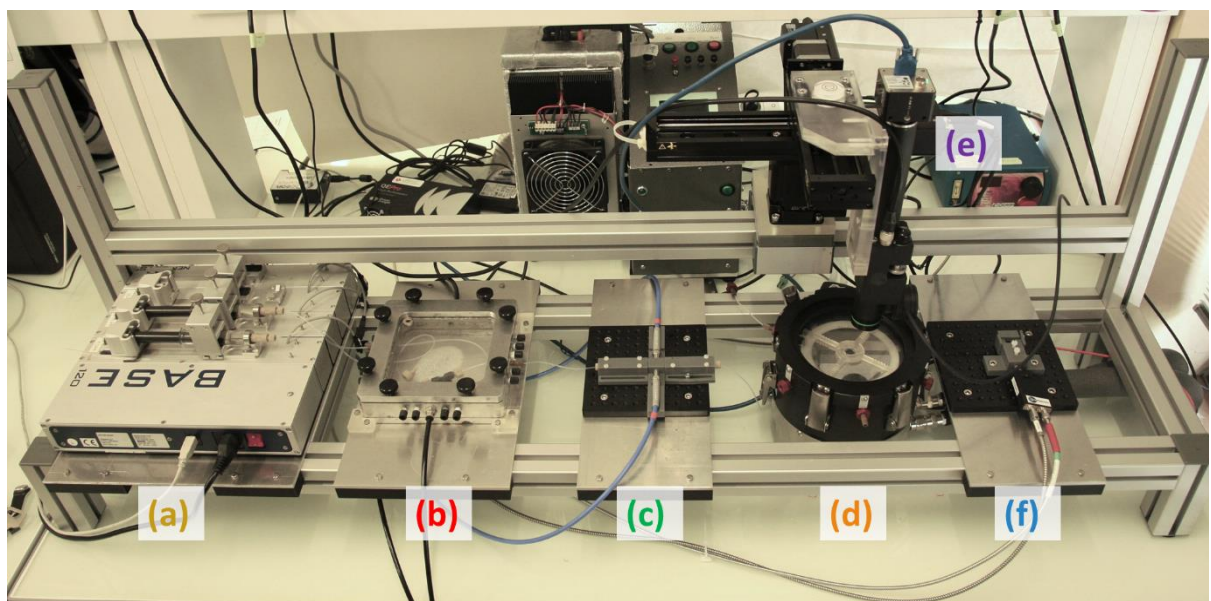

*Figure 1 : Microfluidic platform: (a) Syringes and pumps; (b) Solutions and droplet generation with temperature control; (c) UV characterisation; (d) Droplet storage and cooling; (e) Optical characterisation; (f) Raman characterisation*

Parts of (c), (d) and (f) were designed and manufactured by 3D printing (Formlabs). Different resins were used depending on the function: "High Temp Resin" for the spiral holder in the thermostatic bath (d), "Grey Resin" for UV (c) and Raman holders (f) and "Clear Resin" for the structure.

In (b), the microfluidic circuit is composed of HPLC consumable material (IDEX Health and Science) (SIF 2): junctions and fittings are in polyether ether ketone (PEEK) and tubing (1 mm inner diameter) is in perfluoroalkoxy (PFA). A 7.5cm-long UHPLC stainless steel column of 2.1mm inner diameter and 0.5µm filters contains the powder (ref 5030IP-04021-0075-05 in SIF 2). These polymers are chemically resistant to most solvents, which enables universal use of the platform, whatever the solvent, at a wide range of temperatures.

#### 2.2.1. Syringes and pumps

Solutions, solvents and oil are loaded in glass syringes (gastight syringes series 1000, termination ChemSeal, 5mL and 10mL, Hamilton). The syringes are placed on precision syringe pumps (NeMESYS S, Cetoni GmbH) ((a) in Figure 1) programmed by a computer interface which accurately controls the flow rates.

#### 2.2.2. Saturated solutions and droplet generation

The module generating saturated solutions and droplets consists of a tank filled with thermoregulated water ((b) of Figure 1). The tank contains the UHPLC column (IDEX Health and Science) filled with the API powder (from 50 mg to 100 mg). The solvent flows through the powder bed to generate a saturated solution directly by dissolution of the powder at a given temperature (Figure 2). After the column, the saturated solution flow crosses, in a T-junction, either a pure solvent flow to be diluted for solubility measurements (Figure 2a) or an oil flow for droplet generation (Figure 2b), as described in section 3.

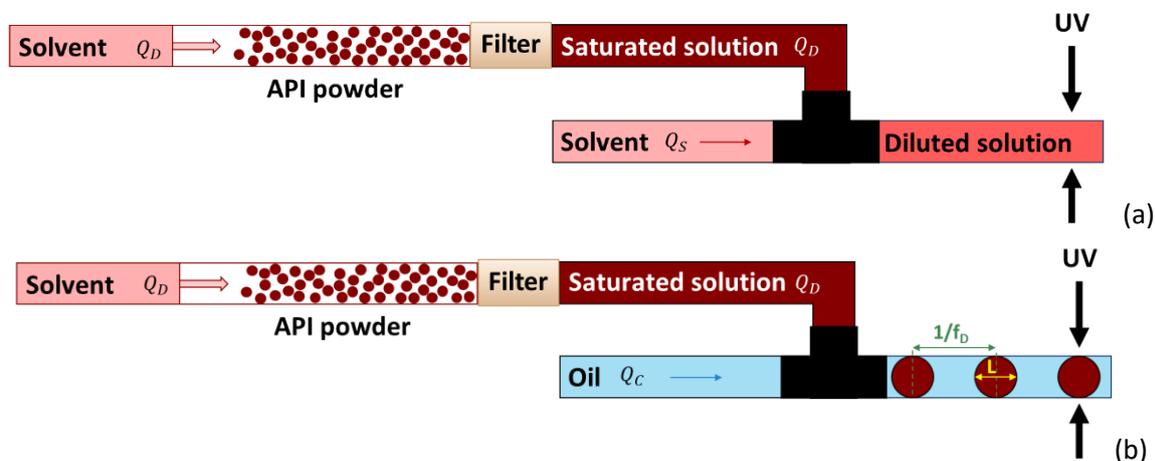

Figure 2: Saturated solution generation followed by (a) dilution for solubility measurements; (b) droplet generation in oil, $Q_D$, $Q_S$ and $Q_C$ are respectively the discontinuous form flow (saturated solution), the dilution solvent flow and the continuous form flow (oil). L and $f_D$ are respectively the length of the droplets and their frequency.

### 2.2.3. Ultraviolet characterisation.

Ultraviolet (UV) characterisation is used for both solutions and droplets, for size and frequency measurements, as described in section 3. UV light is generated by a deuterium lamp (DT-MINI-2-GS, Ocean Insight), emitted and collected through solarised optical fibres, and analysed by a UV-Visible spectrometer (USB2000+, Ocean Insight). For solubility measurements, the solution flows through a clear fused quartz capillary (1mm ID x 1.5mm OD, VitroCom). A 3D-printed resin support was designed in-house to ensure optical alignment of both the optical fibres and the tubing ((c) in Figure 1 and SIF Figure 2).

### 2.2.4. Droplet storage and cooling

Saturated droplets are stored in a temperature-controlled water tank, equipped with a glass window for observation ((d) in Figure 1). The tank contains two independent water circuits. The first is closed and regulated by a thermostatic bath (DYNEO™ DD-600F, Julabo ™) used to regulate the water temperature and perform slow cooling. The second circuit is composed of a homemade thermostatic bath used to rapidly fill the tank chamber with water (in less than 5 minutes). By fast-cooling the droplets, supersaturation is generated at a higher rate.

In order to reduce the observation zone while increasing the number of droplets, the tubing containing droplets is rolled onto a 3D-printed spiral support. This support includes a hard-drive disk platter as reflecting surface for optical microscopy observations. The whole is directly immersed in the tank water, enabling temperature regulation and facilitating optical characterisation due to the matching of PFA tubing and water optical indexes.

### 2.2.5. Dedicated motorised optical microscope

The tubing containing droplets is optically characterised in the tank during cooling. An optical device (Zoom 70 with stepper-motor, TV-Objective 3.1xWD=77mm, L=10mm, TV-Tube 1.0xD=35 mm L=146.5mm, OPTO) and a camera (Exo267MU3, SVS-Vistek) are placed on a motorised XYZ stage (X-VSR40A-PTB2 and XY-LSQ150A, Zaber) to perform a complete screening of the spiral ((e) in Figure 1). Software created in-house (LabVIEW) is used to program XYZ displacements by defining the steps between pictures, the region of interest and time between cycles. A Python code (available on request) is then used to reconstruct a full image of the spiral (Figure 3). Moreover, since the camera and the XYZ stage are mounted on a mobile slide on the MiniTec aluminium structure, they can be displaced to observe different modules at will.

### 2.2.6. Raman spectroscopy characterisation

Droplets are analysed directly in the tubing (in-situ) after the cooling process (off-line) by Raman spectroscopy using a 785nm laser (adjustable output power 350mW, ref. I0785MM0350MF, IPS) connected to a Raman spectrometer (QE Pro+, Ocean Insight) by a probe made of optical fibres (RPB-785-FF-XR enhanced, InPhotonics and OceanInsight) ((f) of Figure 1). A holder was designed and 3D-printed with grey resin to align the probe with the tubing (SIF Figure 2). The distance from the probe lens to the tubing, as well as the tubing vertical position, can easily be adjusted to focus the laser beam at the best position inside the droplet and optimise the signal. The laser beam position can be controlled by the optical characterisation module (2.2.5) thanks to a reflecting surface under the tubing. Lastly, the measured Raman spectra are analysed with Python.

### 2.3. Interfacial energy measurement

The interfacial energies are measured by the pendent drop method using an OCA-20 device (Dataphysics) and the SCA20_U software. A pendent drop of the densest media in a vertical 1mL syringe is generated inside the less dense media contained in a transparent cell. The drop is subjected to gravitational effects and interfacial energy on the aperture of the tip of the syringe (Nordson, external diameter 0.82mm).

### 2.4. Supplementary solubility measurements

For verification purposes, solubility measurements were performed in thermostatted millivials. A 1mL glass vessel is placed in a Monopuits (Anacrismat) Peltier thermostatted cell and observed under an optical microscope (Nikon, objective x4). The set-up was previously described by Boistelle et al.[23]

## 3. Methods
### 3.1. Interfacial energy measurement

The interfacial energy was measured by the pendent drop method, first described by Tate [24]. The drop is generated using a low flow of liquid and pictures are taken regularly, until the drop detaches and falls. The last picture before detachment, corresponding to the balance between interfacial energy and gravitational effects, is analysed by the software (SCA20) to detect the shape of the drop at equilibrium and the curvature radii. These characteristics are related to the interface energy by the Laplace-Young equation, yielding the interface energy between two fluids, e.g. oil/air, solvent/air and oil/solvent. For each pair of fluids, the measurement is performed on at least 10 drops. However, interfacial energy being highly sensitive to the environment (namely temperature and presence of microdust), the results presented in 0 must be considered with caution.

### 3.2. Solubility measurement

#### 3.2.1. Microfluidic set-up

In our microfluidic set-up, solubility was measured at different temperatures for stable and metastable forms using the step dilution method developed by Peybernès et al [17]. However, special care is required with metastable forms or in the case of solvate formation, which may induce phase transition. When several measurements are realised with the same powder, each should be performed at a higher temperature than the last, in order to avoid recrystallisation (in the case of direct solubility). A back-pressure regulator (IDEX Health and Science) is added at the outlet of the set-up to prevent solvent cavitation.

### 3.2.2. Thermostatted millivials

In thermostatted millivials, supplementary measurements were performed at larger scale via a method similar to that described by Detoisien [25]. Suspensions are produced by adding a known amount of powder to a given volume of solvent, in 1mL vials. The vials are then slowly and incrementally heated (by increasing the temperature every few hours) while being observed in a Monopuits (Anacrismat) Peltier thermostatted cell [26]. Complete dissolution of the crystals occurs at a temperature corresponding to the equilibrium temperature at the starting concentration.

### 3.3. Droplet generation

As represented in Figure 2b, the droplets are generated by making a flow of oil, the continuous fluid here (rate $Q_C$), cross a flow of saturated solution, the dispersed fluid (rate $Q_D$). As described in section 2.2.2., the saturated solution is generated from powder. The droplets' length (L) and frequency ($f_D$) are then characterised by UV spectroscopy.

Different droplet generation regimes can be defined depending on the capillary number value Ca, which is defined by (1).

$$Ca = \frac{\mu_C \times v_C}{\gamma_{CD}} \quad (1)$$

$\mu_C$ is the dynamic viscosity of the continuous fluid, $v_C$ the velocity of the continuous fluid and $\gamma_{CD}$ the interfacial energy between the two fluids, measured by the pendant drop method (Dataphysics, OCA-20). Since our aim was to produce quasi-spherical droplets of regular size and frequency, with a diameter close to the tubing diameter (here 1mm) to reduce droplet mobility, and thus coalescence (as no surfactant is used), the squeezing regime (Ca < 0.01) was the most appropriate. Zhang et al [27]. proposed a model to link size and frequency of droplets to Ca. Here, we used flows between 0.2µL/s and 1.5µL/s to obtain low Ca and regular droplets. Droplet size and frequency were analysed by UV-spectroscopy (3.4 and SIF 3).

### 3.4. UV-spectroscopy characterisation

For ultraviolet characterisation, the integration time was chosen from a range of between 10ms and 100ms. The wavelength was chosen depending on the maximum absorbance of the solution. For Rimonabant in acetonitrile or ethanol and Irbesartan in ethanol, intensity was integrated from 250 to 300nm. For Sulfathiazole in acetonitrile and water, intensity was integrated from 200 to 300nm. For Aripiprazole in acetonitrile, the intensity was integrated between 220 and 300nm.

Since absorbance depends on temperature, the UV cell must be at room temperature for solubility or concentration measurements. However, reducing the temperature increases the risk of recrystallisation in the solution. In the case of a flow, dilution is used to prevent the product crystallising. The known dilution rate is used to determine the concentration.

### 3.5. Droplet storage and cooling

The tank containing the droplet storage tubing is preheated to at least 5°C higher than the generation temperature to dissolve any crystal which may have appeared between the generation and the storage areas. Different cooling profiles can be used: from slow cooling ramps at -0.4°C/h up to fast cooling from 65°C to 10°C in less than 5 min. Here, only ramps at -0.5°C/h and -5°C/h were used (cf. part 4.2).

*3.6. Optical characterisation*

Since the observation field of the optical device is too small to observe the entire droplet-containing tubing, the motorised XYZ stage enables screening to be performed in the three directions. A matrix of 180 camera positions is defined. The resulting pictures are then combined on Python to reconstruct a global image of the tubing (Figure 3a).

For crystallisation screenings, cycles of 180 pictures are performed every 30 min to 2h during cooling and storage. For nucleation statistics, the nucleation rate is monitored over time and temperature by counting droplets containing crystals and empty droplets for the different picture cycles. Crystal habit can indicate polymorphism, however it is not sufficient evidence, and the different crystal habits are further analysed using Raman spectroscopy or XRD.

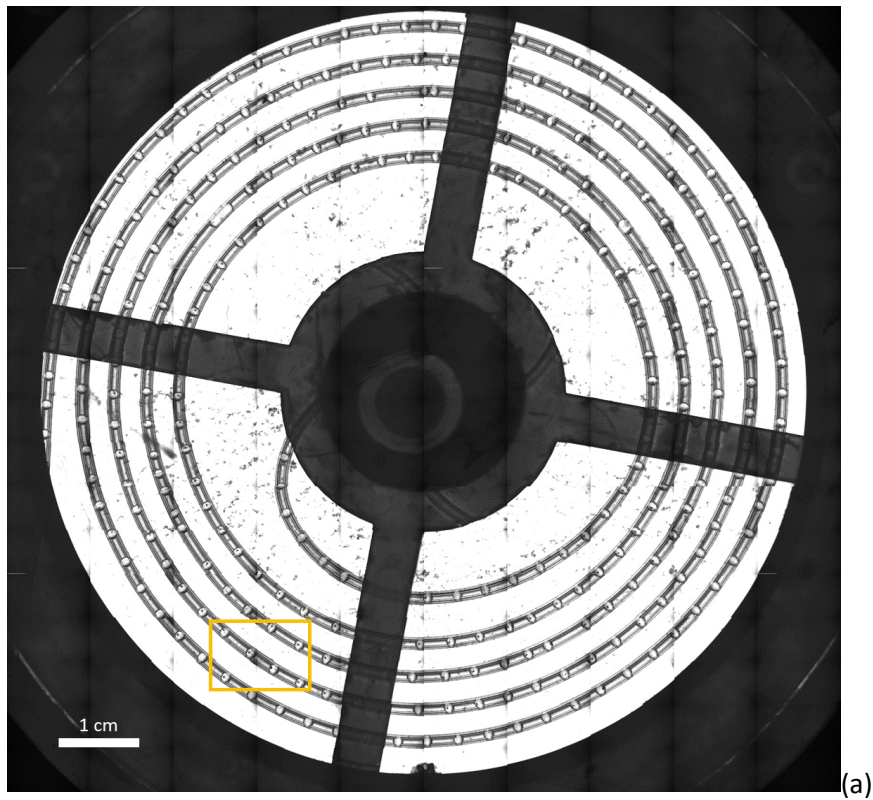

(a)

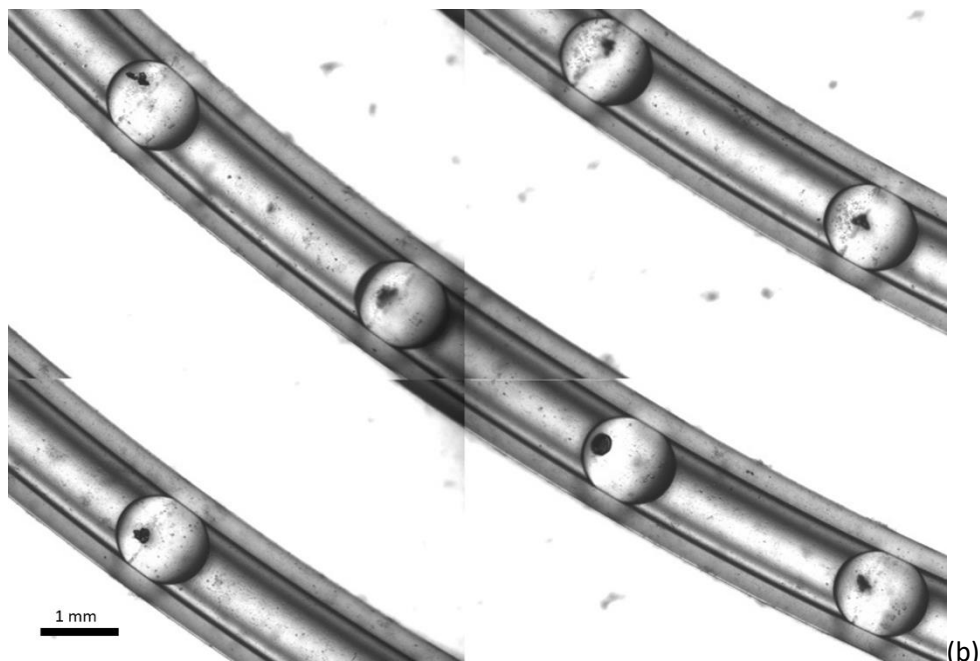

*Figure 3: Sulfathiazole in water droplets generated at 70°C in GPL107 oil and cooled to 10°C, tubing of 1mm inner diameter: (a): full spiral reconstruction from 180 pictures; (b) zoom on (a) at the junction of four pictures, zone indicated by the square in (a).*

### 3.7. In-situ Raman characterisation

Each droplet is analysed directly in the tubing. The integration time is set at 100 to 500ms. For each molecule, one or several zones of interest are defined. These zones should include at least a peak corresponding to the crystallised product, and avoid the signals of the tubing, the oil and the solvent (SIF 6). In SIF 6, such a peak can be found near 1650cm$^{-1}$ for Rimonabant.

The Raman spectra of each droplet is represented in a "colormap" as shown in Figure 4a. The first step is identifying droplets containing crystals. A Raman Shift range, here between 1630 and 1730cm$^{-1}$, is chosen. In Figure 4b, the signal intensity is normalised by the maximum intensity of this zone. If the droplet is empty, the normalised signal then has an intensity close to 1 over the whole zone. In contrast, if the droplet contains a crystal, the normalised intensity is 1 at the peak and close to 0 around the peak. Thus, a Raman shift is chosen outside the peak, e.g. at 1640cm$^{-1}$ in Figure 4b where the normalised intensity is compared to a threshold value (here 0.7) to define whether or not the droplet contains a crystal. In Figure 4b, droplets containing a crystal are marked by a black dot at 1640cm$^{-1}$.

Next, crystalline forms in droplets containing crystals are determined by comparing peak positions in one or several zones. In Figure 4c, Form I and Form II of Rimonabant are distinguished at between 1630 and 1730cm$^{-1}$. Spectra with a peak below 1670cm$^{-1}$ are Form I and the others are Form II.

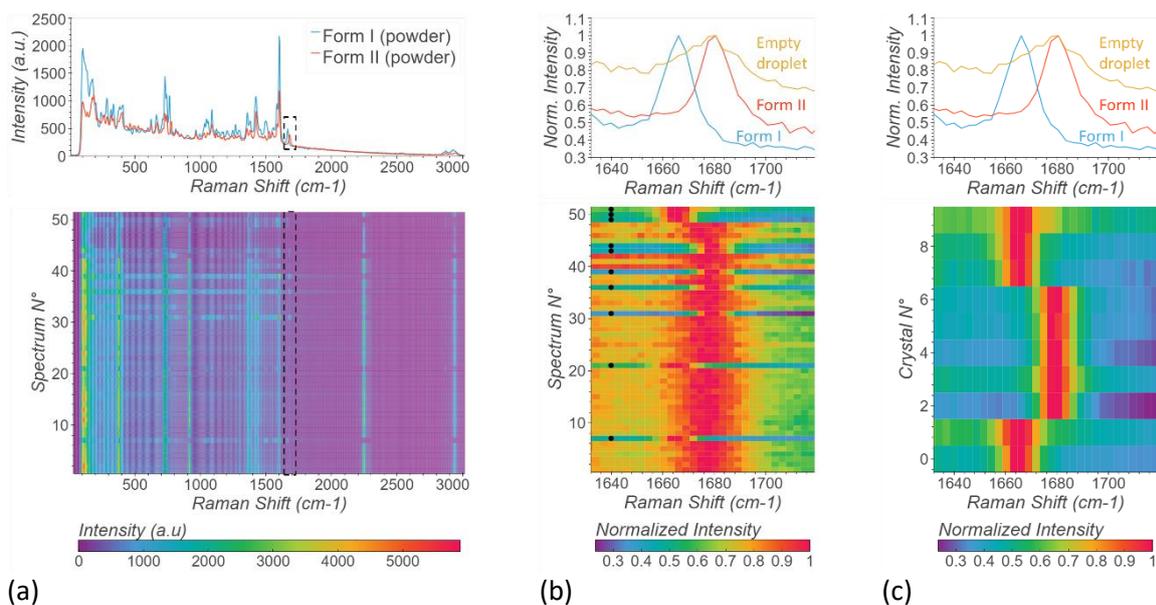

*Figure 4 : Raman spectra of Rimonabant droplets in acetonitrile, generated at 60°C and cooled to 10°C: (a) top: reference spectra of Rimonabant Forms I and II from powder; bottom: full representation of 145 droplets; (b) and (c) top: normalised reference spectra of Rimonabant Forms I and II from powder and an empty droplet between 1630 and 1730cm$^{-1}$ (b) bottom: normalised representation of 51 droplets between 1630 and 1730cm$^{-1}$ for the selection of droplets with crystals. The black circles indicate the droplets containing crystals; (c) bottom: normalised representation of 10 droplets containing crystals between 1630 and 1730cm$^{-1}$ for the form analysis.*

For the interpretation, the spectrum can be compared to references. Raman analysis can also be supplemented by XRD after extracting some crystals from the droplets. We simplified the method described by Gerard et al. [28], since we manipulate organic molecules, which do not need to be cryogenised and can be manipulated in ambient conditions and outside the solution. Here, the references for Rimonabant were measured from powder of form I and form II. For Sulfathiazole, the spectra were compared with the literature [19], [20], [29].

Crystals can be analysed directly by Raman spectroscopy in-situ, and several crystalline forms can be identified. But when crystals are small or located at the droplet/oil interface, the signals of oil or solvent can be too intense compared to the crystal signal, making it difficult to use the whole spectrum. In this case, special care must be taken in choosing the zones for form identification.

However, under some conditions, several crystals can be analysed in a single droplet, as illustrated in Figure 5. In this example, the droplet contains two crystals with different habits (Figure 5a). Here, the laser beam was focused on each crystal (Figure 5b and c) to obtain two different Raman spectra (Figure 5d): the left crystal appeared to be Form II of Rimonabant while the other was form I.

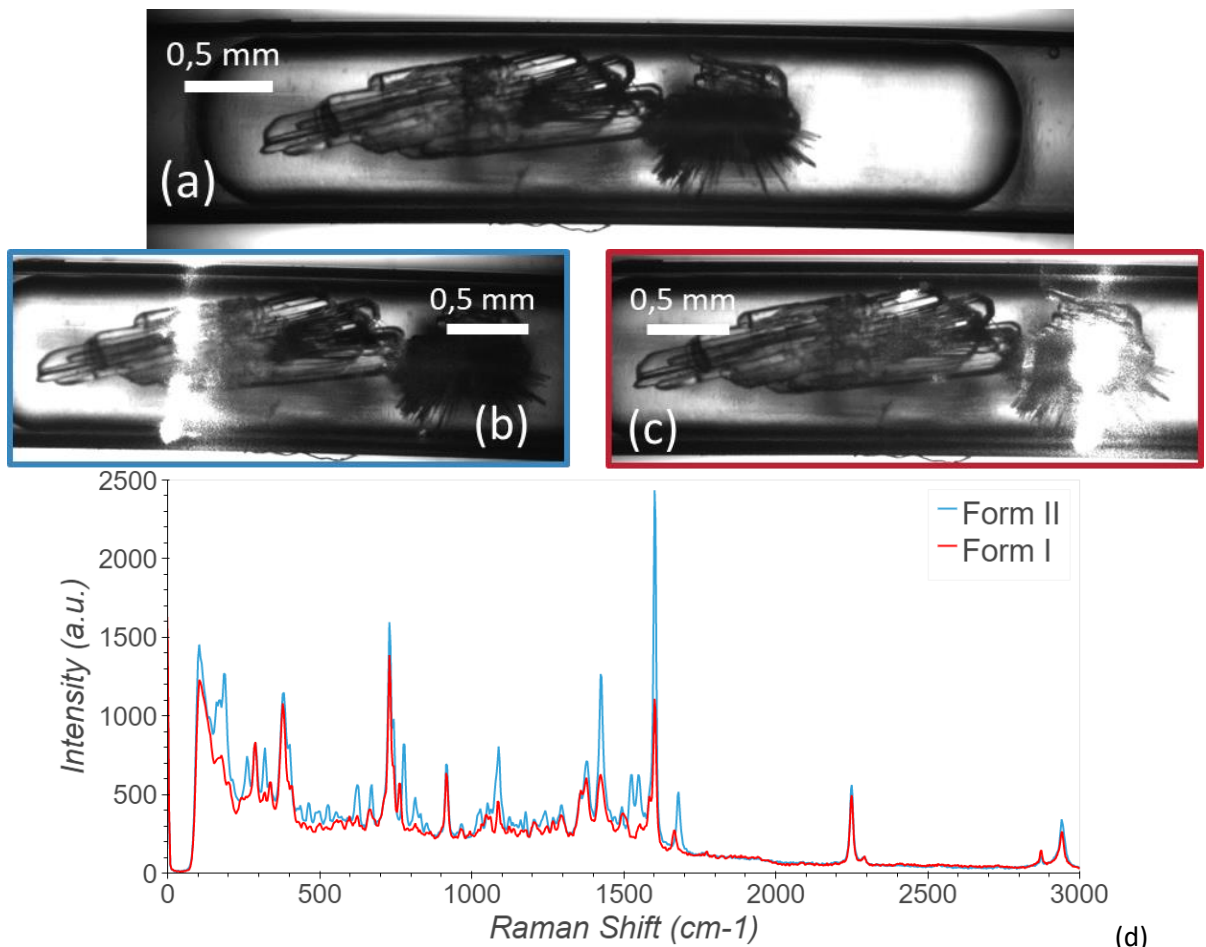

*Figure 5: (a) Plug of acetonitrile containing Rimonabant crystals, generated from saturated solution at 70°C and cooled to 30°C; (b) Laser focused on the left crystal; (c) Laser focused on the right crystal; (d) Raman spectra corresponding to the left crystal (Form II) and the right crystal (Form I).*

## 4. Results and discussion

### 4.1. Solubility measurements

Solubilities of Irbesartan, Rimonabant and Aripiprazole were measured by the microfluidic method described in part 3.2.1, with 5% error. The range of temperatures that can be explored with this method is at least 15 to 70°C, and concentrations up to 100mg/mL can be measured (SIF 7). Measurement of each value of solubility at a given temperature took 1h, including heating time and concentration stabilisation. For the lowest solubilities, it was possible to use a full column of powder (60 to 90mg) for the entire solubility curve (eg. Aripiprazole) or at least for several points. However, for the highest solubilities (eg. Rimonabant above 40°C), a full column of powder was needed for a single value of solubility.

#### 4.1.1. Irbesartan

Figure 6a compares measured solubilities of Irbesartan form A and form B in ethanol to the solubility curve obtained on form A by Wang et al. [30]. Form B is the most soluble desmotrope [22] in ethanol. This result shows that our method can be used to measure the solubility of more soluble forms, i.e., metastable forms in the case of strict polymorphism, provided that no phase transition occurs during the measurement. When XRD spectra of the powders in the column were measured before and after the experiments to identify a potential phase transition, none was observed.

*4.1.2. Rimonabant*

Solubility curves of Rimonabant form I and II in acetonitrile, measured by our microfluidic method are represented in Figure 6b and c. We compare them with our measurements in thermostatted millivials to validate the method in Figure 6c. Both polymorphs present similar solubilities. Hence, they are impossible to differentiate by the usual methods, as already observed in different solvents by Fours [31] and Alcade et al. [32] (Figure 6b). This result shows that it is difficult to determine their relative stability based on solubility. Fours [31] reported an enantiotropic system with form I stable under a transition temperature between 50 and 58°C, while Perrin et al. [18] established a pressure-temperature diagram showing a monotropic system with stable form II.

*4.1.3. Aripiprazole*

The starting powder of Aripiprazole was a mixture of forms II and III. Form II is more stable than form III, but it is not the most stable form of the system at room temperature [33]. Therefore, a preliminary experiment was performed in microfluidics at 20°C to highlight a possible phase transition. Acetonitrile was injected through the powder in the column at a constant flow rate of 0.30μL/s and the saturated solution was diluted by an acetonitrile flow rate of 0.34μL/s. The measured UV signal is represented in SIF 8. The intensity maxima at the beginning of the experiment correspond to air bubbles resulting from the wetting of the powder. Then we observed an unstable signal of a low intensity, meaning high concentration, for the first hour. This can be interpreted as the dissolution of the metastable polymorph (here form III), and the intensity drop as its recrystallisation into form II. After 1h however, the signal reached a plateau, indicating no further phase transition. At the end of the experiment, the XRD analysis of the resulting powder showed that only pure form II remained in the column. This result illustrates that this set-up can be used to purify a mixture of several polymorphs to a more stable form and measure its solubility.

Thus, the solubility curves of Aripiprazole form II measured in acetonitrile from a mixture of forms II and III, with microfluidics, are consistent with measurements in thermostatted millivials (Figure 6d).

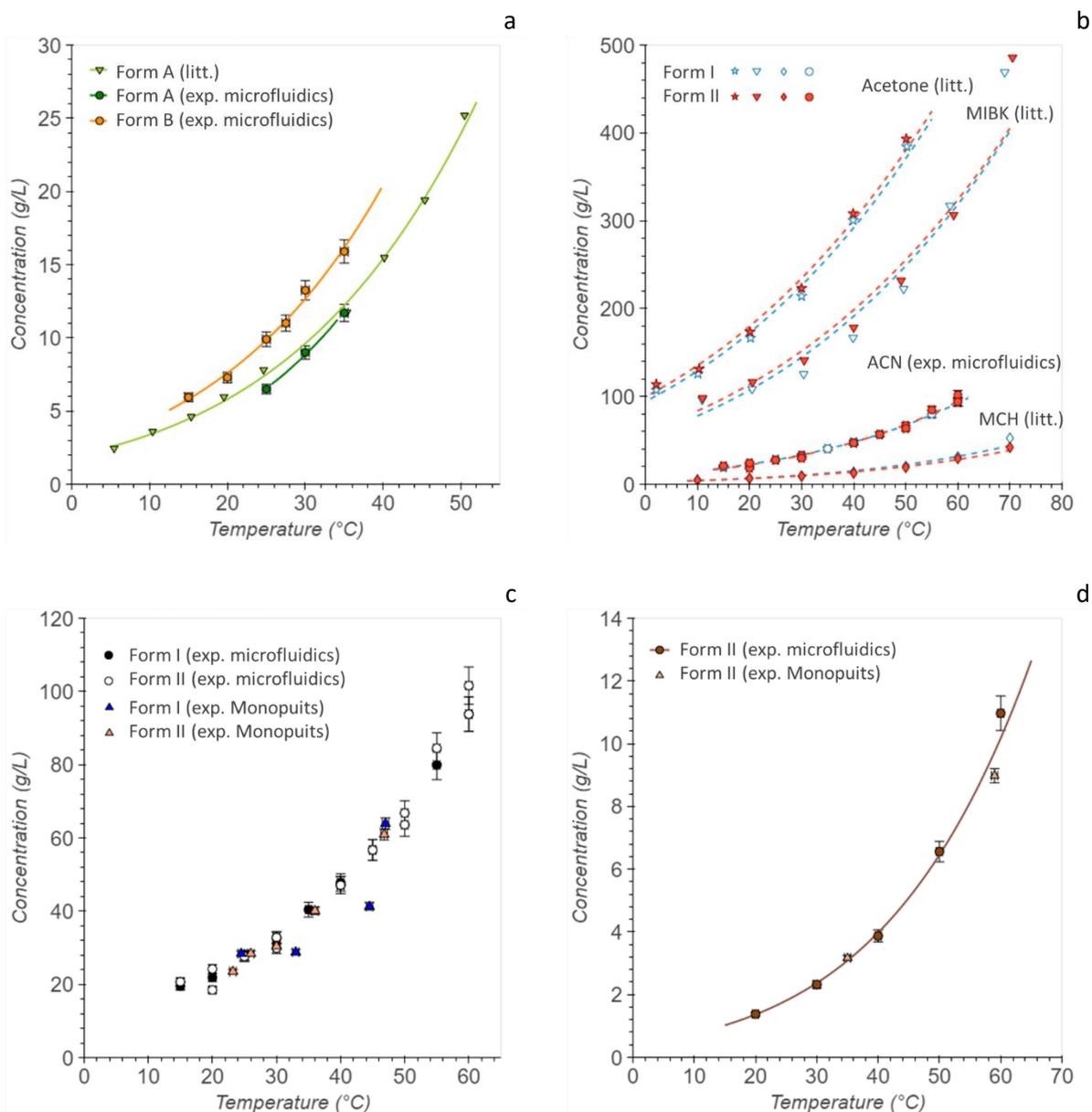

*Figure 6: (a) Solubilities of Irbesartan forms A and B measured in ethanol with microfluidics. Solubility of form A is compared with the literature [30]. (b) Solubilities of Rimonabant forms I and II measured in acetonitrile with microfluidics and compared with solubilities of forms I and II from the literature in acetone, methyl isobutyl ketone (MIBK) and methylcyclohexane (MCH) [31], [32]. (c) Solubilities of Rimonabant forms I and II measured in acetonitrile with microfluidics and Monopuits. (d) Solubility of Aripiprazole form II measured in acetonitrile with microfluidics and thermostatted millivials (Monopuits). Data are fitted with Van't Hoff equation..*

### 4.2. Statistics for crystallisation in droplets

We generated saturated droplets at 70°C and cooled them down to 10°C using different cooling profiles, maintaining them at this temperature for several hours or days. Pictures were taken every 30 min to 2h as described in part 2.2.5. In addition to polymorph screening, these experiments provide statistics on crystalline form nucleation in droplet-based microfluidics. Droplets containing crystals and empty droplets were counted on several picture reconstructions. The example shown in SIF 9 is the last picture cycle of a screening on Sulfathiazole in water (-5°C/h ramp). In the experiments presented here, no solution-mediated phase transition was observed.

#### 4.2.1. Sulfathiazole in water

For Sulfathiazole in water, we compared two different ramps of temperature from 70°C to 10°C:

- The first ramp had a cooling rate of -5°C/h and the droplets were maintained at 10°C for 3 days after the end of the ramp. The percentage of droplets containing crystals as a function of time is plotted in Figure 7a in parallel with the temperature profile. The first crystal appeared during the cooling, around 29°C. A few more crystals (around 5%) appeared before the temperature reached 10°C, followed by a plateau where no more crystals appeared for a few hours. Then the number of crystals increased again. Counting stopped after 2.5 days at 10°C, at which point we observed 51% of droplets containing crystals (162 crystals/319 droplets).

- The second ramp had a slower cooling rate of -0.5°C/h and the droplets were maintained at 10°C for almost one day. As in the previous experiment, the first crystal appeared around 30°C. The percentage of crystals first reached a plateau of 5% (Figure 7b). However, this plateau occurred during the cooling, between 25°C and 19°C. A second wave of nucleation occurred between 19°C and 17°C, reaching more than 90% nucleation. Then the number of crystals increased slowly, reaching 95% at a temperature of 10°C. Finally, after almost one day at 10°C, 100% of the 453 droplets contained crystals.

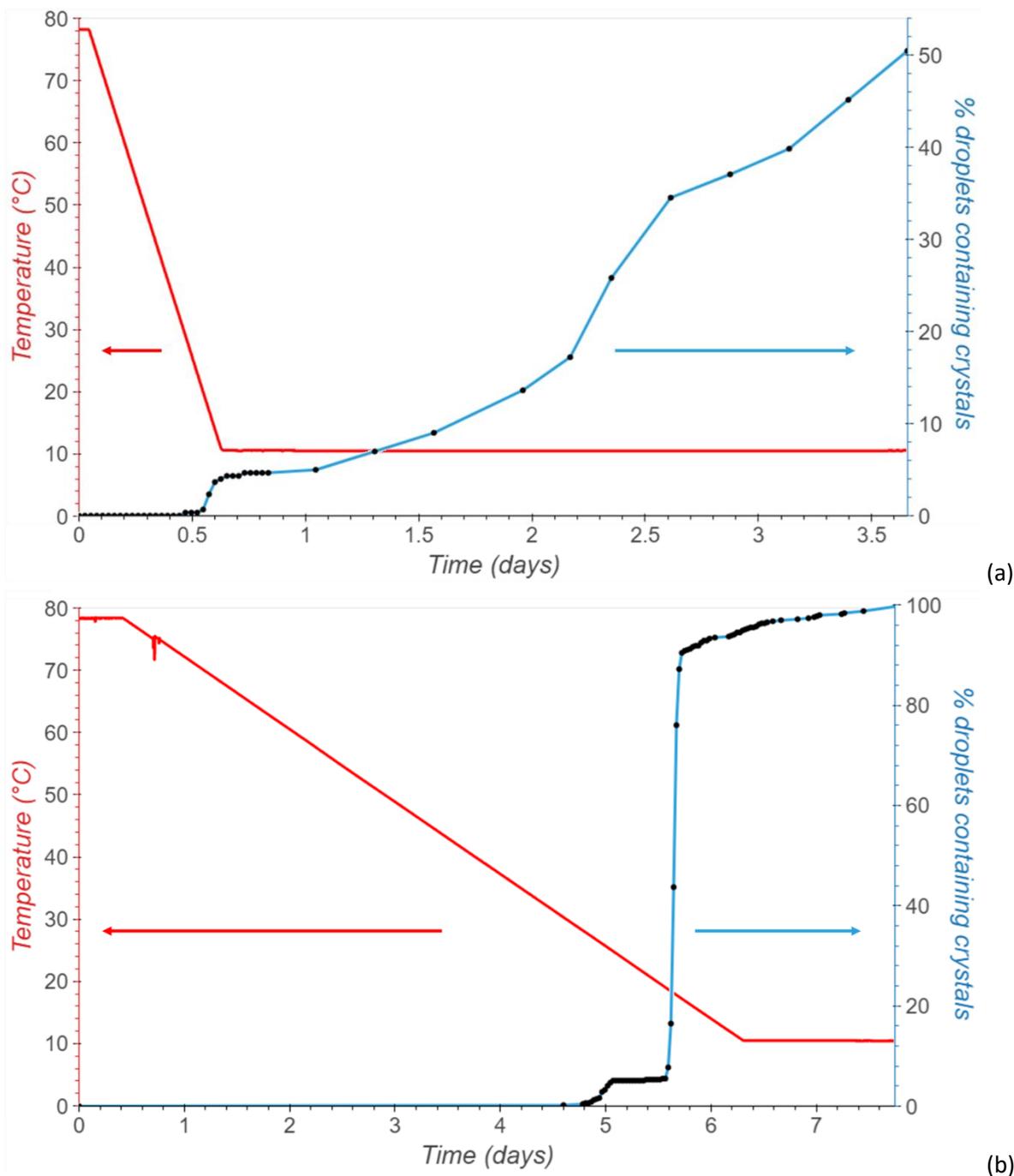

Figure 7: Percentage of droplets of water containing crystals of Sulfathiazole as a function of time (blue) and temperature profile as a function of time (red).(a) for a fast-cooling ramp (-5°C/h); (b) for a slow cooling ramp (-0.5°C/h)

### 4.2.2. Sulfathiazole in acetonitrile

For Sulfathiazole in acetonitrile, we compared two different temperature profiles from 70°C to 10°C:

- The first ramp had a cooling rate of -5°C/h ramp and the droplets were maintained at 10°C for 11 days. At the end of the experiment, only 19% (89/465 droplets) of the droplets contained crystals. Most of the crystals appeared during the cooling. Since solubility is five times higher in acetonitrile than in water [29], a higher nucleation rate could be expected [34]. However, the increase in supersaturation from 70°C to 10°C is almost double in water compared to acetonitrile. Thus, the poor nucleation rate we observe illustrates the influence of supersaturation on nucleation, which is in agreement with Peybernès et al [29].

- The second temperature profile consisted of a quench from 70°C to 10°C and the droplets were maintained at 10°C for 6 days. No crystal appeared during cooling, nor at 10°C.

In all of these experiments concerning sulfathiazole, we observe that the slower the cooling rate, the higher the number of crystals.

### *4.2.3. Rimonabant in different solvents*

Similar experiments were performed with Rimonabant in acetonitrile, ethanol and ethanol-water mixtures with different temperature profiles. However, the nucleation rates were too low, producing no crystals in some cases, even after several days at 10°C. These results underline the limitations of volume reduction, which may considerably increase nucleation time, as previously described by Hammadi et al.[35] and observed by Teychené et al. [36] on various organic molecules in organic solvents.

### *4.3. Polymorph screening and statistics of nucleated forms*

Thanks to the versatility of our set-up, the solid can be directly analysed in the tubing (in-situ) after the cooling process (off-line) by Raman spectroscopy. Thus, for the polymorphism screening, we used the experiments and the results described in 4.2. We studied Sulfathiazole in water with two different ramps (-5°C/h and -0.5°C/h) and in acetonitrile with one ramp (-5°C/h), by Raman and XRD analysis on crystallised droplets.

### *4.3.1. Sulfathiazole in water*
#### *4.3.1.1. Fast ramp (-5°C/h)*

For Sulfathiazole in water, after the -5°C/h ramp from 70°C to 10°C and 3 days at 10°C, 319 droplets were analysed by Raman spectroscopy, among which 162 droplets were crystallised. The forms obtained are summarised in Figure 8a:

- Form II and form IV appeared respectively in 1.85% and 11.7% of the droplets. They did not show any representative habit (Figure 8b) and could not be optically distinguished.

- The majority of the crystals (85.8%) appeared to be an unknown form, called form U1 here, presenting a particular habit (Figure 8c). This habit was previously observed by Peybernès [37] in a few droplets of a preliminary experiment but no XRD was performed at that time. Here, the XRD pattern is provided in supplementary information (SIF 5). This pattern does not correspond to any known form reported in the Cambridge Structural Database. In order to study the unknown form U1, we performed an additional dissolution experiment, presented in supplementary information (SIF 12). It shows that form U1 has higher solubility than form IV, and thus than form II, whose solubility is similar to form IV [19].

- A last crystal, called form U2 here, presents a slightly different Raman spectrum (SIF 11). However, due to its needle-like shape and small size (Figure 8d), we did not manage to collect the crystal for further analysis (e.g. XRD) in this first experiment.

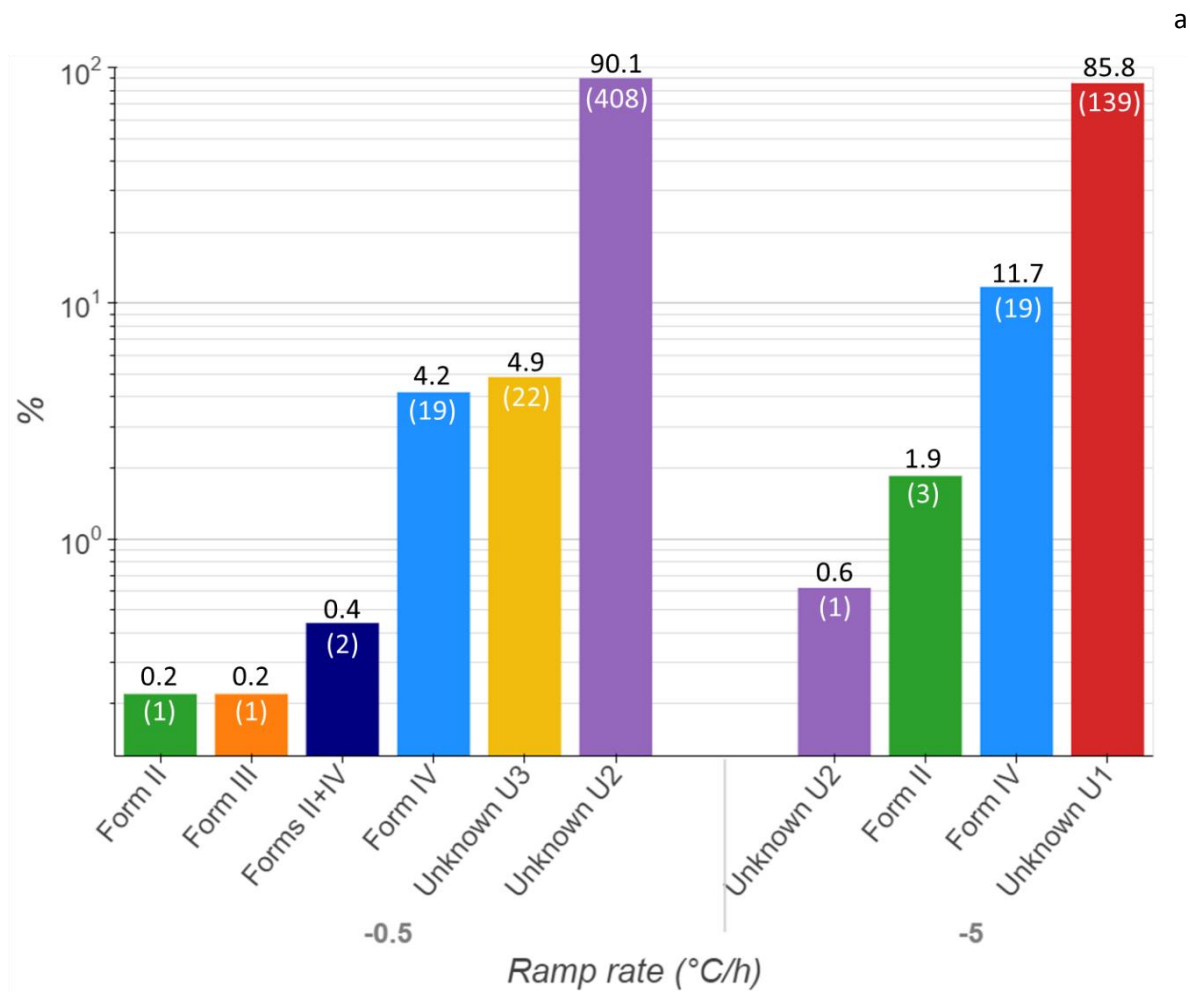

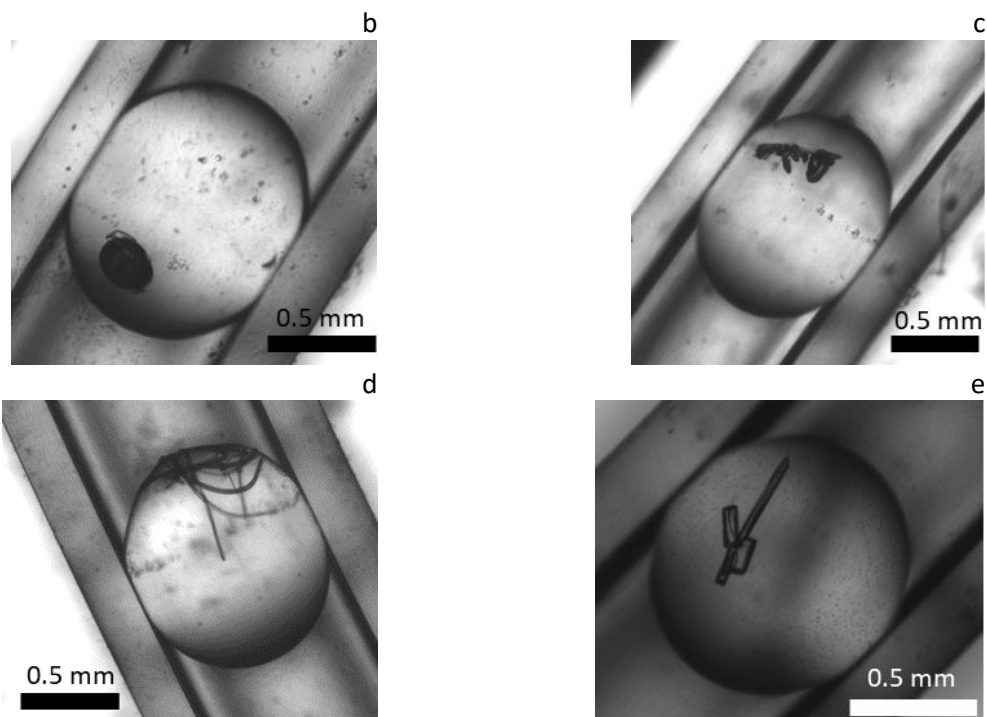

*Figure 8: (a) Percentage (and number of crystals) of Sulfathiazole forms II, III, IV, mixture II+IV and unknown U1, U2 and U3 among the droplets containing crystals with -0.5°C/h and -5°C/h ramps in water from 70°C to 10°C. ( b, c, d and e) Crystals habits of Sulfathiazole obtained in water, mainly for (b) forms II and IV; (c) unknown U1; (d) unknown U2; (e) unknown U3.*

The form statistics presented in Figure 8a are consistent with the stability order established by Munroe et al. [19], with form II being more stable than form IV. We additionally report a new metastable form, U1. Furthermore, we also clearly observed a kinetic effect on crystallisation which competes with the thermodynamic effect: the first crystals to nucleate appear to be forms II and IV, which are thermodynamically more stable. In contrast, the metastable form U1 appears several hours or days later, while the temperature is maintained at 10°C.During the cooling process, the supersaturation increases while the temperature decreases. The experimental conditions inside each droplet move on the phase diagram from undersaturated or saturated to supersaturated. They first cross the solubility curves of forms II and IV respectively, before crossing the solubility curve of form U1. The same trend is found for the metastable limit of the three forms. This may explain why at the beginning of nucleation, the more stable forms (II and IV) nucleate: they show higher supersaturation. At that moment, the solution is above the metastable limit for forms II and IV but inside the metastable zone for form U1. However, after a waiting time, kinetics takes the lead over thermodynamics, leading to the nucleation of the metastable form U1 according to Ostwald's rule of stages.

### 4.3.1.2. Slow ramp (-0.5°C/h)

For the second experiment with Sulfathiazole in water, after the -0.5°C/h ramp from 70°C to 10°C and a day at 10°C, we analysed 453 crystallised droplets by Raman spectroscopy. The forms obtained are summarised in Figure 8a:

- Form II and form III each nucleated in a separate single droplet (0.221%), while 4.19% of nucleated droplets were form IV. Two crystals appeared to be a mixture of forms II and IV (SIF 11). However, none of these forms showed any representative habit (similar to Figure 8b), so they could not be optically distinguished.

- The majority of the crystals (90.1%) appeared to be the unknown form U2 described in 4.3.1.1, according to the Raman spectra (SIF 11). The crystals showed the same habit as in Figure 8d. In this experiment (compared to 4.3.1.1), we managed to collect the crystals, and the XRD pattern is provided in supplementary information (SIF 5). This pattern does not correspond to any known form reported in the Cambridge Structural Database.

- Another unknown form (here called U3) accounted for 4.86% of the crystals. The Raman spectrum does not correspond to any reference (SIF 11), and nor does the XRD pattern (SIF 5). It showed a well-faceted habit, as represented in Figure 8e.

Similarly to the previous experiment (in 4.3.1.1), the known forms II, III and IV nucleated earlier than the most common forms U2 and U3, which appeared during the second wave of nucleation, described in 4.2.1. This may also be explained by the competition between thermodynamics and kinetics.

### 4.3.2. *Sulfathiazole in acetonitrile*

For Sulfathiazole in acetonitrile, after the cooling ramp of -5°C/h and the 11-days plateau at 10°C, the 89 crystallised droplets displayed crystals with a wide range of habits as shown on SIF 13. Moreover, they were larger than those obtained in water, as expected from the higher solubility of Sulfathiazole in acetonitrile at 70°C. However, only forms II and IV were produced, according to Raman analysis and as confirmed by XRD on several crystals. Form II was the most frequent polymorph (97%, 86/89 crystals). Only two pure crystals of form IV were obtained (2%). One droplet contained two crystals: one of pure form II, and one showing peaks of both forms II and IV.

These results are consistent with Peybernès [37] who only obtained a small proportion of crystals, all which appeared to be form II, from a slower cooling profile.

5. Conclusions

We have presented a modular microfluidic platform for crystallisation studies developed in our laboratory, and which can easily be implemented in any R&D laboratory. By dissolution of a powder, our set-up generates saturated solutions that can be used for solubility measurements or distributed in microdroplets. The solutions or the droplets are characterised by UV-spectroscopy for solution concentration measurements or to characterise microdroplet sizes and frequencies. Then, microdroplets are stored and cooled using different temperature profiles, and sequential image acquisition is performed during the crystallisation to monitor nucleation. Lastly, crystals in microdroplets are characterised by Raman spectroscopy performed directly in the droplets.

Solubility measurements were performed on different forms of pharmaceutical molecules in several organic and aqueous solvents. The method enables solubilities to be measured for both stable and metastable forms reaching different ranges of concentration and temperature. With the case of Aripiprazole, we also illustrate that this method can be used on a polymorph mixture to generate a pure form, whose solubility can also be measured.

Secondly, we have provided nucleation statistics after cooling. With Sulfathiazole in water and in acetonitrile, we show that the cooling rate influences nucleation. Slower cooling profiles appear to promote nucleation, whereas fast-cooling profiles reduce the number of crystals. We also demonstrate the limitations of microfluidics for crystallisation by cooling: reducing the crystalliser volume considerably increases nucleation induction time. Hence, sufficient supersaturation must be reached to obtain crystals in a reasonable time. Different crystallisation methods have already been explored to reach higher supersaturation, like droplet evaporation [38], [39], addition of a non-solvent, or inducing nucleation with an external field [40]–[43]. Our future research plans include implementing these methods on our modular platform.

In reporting polymorph screening on Sulfathiazole and the statistics for nucleated forms, we have shown in water that the cooling rate influences polymorphism results. We discovered three unknown forms whose XRD patterns and Raman spectra do not match any referenced form. We observed that the known stable forms nucleate earlier than the unknown forms, but that at the end of the experiments, most of the crystal are of unknown forms. This illustrates the competition between thermodynamics and kinetics. In acetonitrile, only known forms were observed, form II being the most common, as already observed by Peybernès [37]. All these findings indicate that our microfluidic platform is a powerful tool for polymorph screening that can be used in the pharmaceutical industry to discover new forms of active pharmaceutical ingredients.

**Associated content**

**Note**

The authors declare no competing financial interest.

**Acknowledgments**

We thank Sanofi R&D for financial support. We thank Marjorie Sweetko for English revision.

**Supplementary Information File**

**SIF 1.    Properties of the GPL107 oil**

| Estimated Useful Temperature Range (°C)[a] | | -30 to 288 |
|---|---|---|
| **Base Oil Viscosity (cSt)[a]** | 20°C | 1535 |
| | 40°C | 450 |
| | 100°C | 42 |
| | 204°C | 6 |
| | 260°C | 3.3 |
| **Oil viscosity index[a]** | | 145 |
| **Oil density (g/mL)[a]** | 0°C | 1.95 |
| | 100°C | 1.78 |
| **Measured surface tension (mN/m)[b]** | Air | 18-20 |
| | Acetonitrile | 13-15 |
| | Ethanol | 9 |
| | Water | 52-55 |

*Table 1 : Properties of fluorinated oil GPL 107. (a) given by Krytox[TM], (b) measured in this study by the pendant drop method at 20°C*

## SIF 2. Microfluidics components

| Reference | Designation | |
|---|---|---|
| 1507L | PFA Tubing Natural 1/16" OD x .040" ID x 50ft | 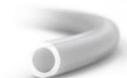 |
| 3160--04-21 | Packing Adapter, IsoBar 2.1mm ID | 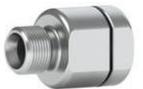 |
| 5030IP-04021-0075-05 | IsoBar Systems 2.1 mm ID, Parker Port 7.5cm | 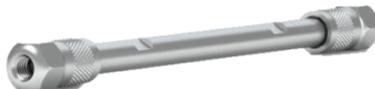 |
| F-120X | One-Piece Fingertight 10-32 Coned, for 1/16" OD Natural - 10 Pack | 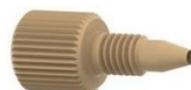 |
| P-200X | Flangeless Ferrule Tefzel™ (ETFE), 1/4-28 Flat-Bottom, for 1/16" OD Blue-10 Pack | 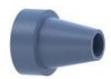 |
| P-287X | Super Flangeless Nut for 1/16" or 1/32" OD Tubing - 10Pack | 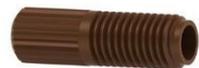 |
| P-316 | Plug for 1/4-28 Flat-Bottom Ports | 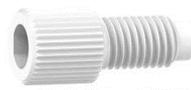 |
| P-703 | Union Assembly PEEK .050 thru hole, for 1/8" OD | 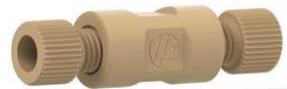 |
| P-728 | PEEK Tee .050 thru hole Hi Pressure, F-300 | 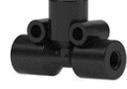 |
| P-733 | PEEK Shut-Off Valve | 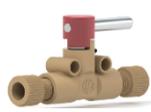 |
| P-760 | PEEK Union for 1/16" OD Tubing | 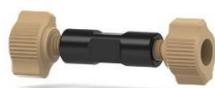 |
| P-790 | Back Pressure Regulator Assembly (5psi) | 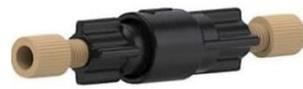 |
| U-467R | Plug for 10-32 Coned Ports | 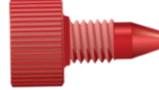 |
| XP-235X | Flangeless Fitting Short, PEEK, 1/4-28 Flat-Bottom for 1/16" OD Natural - 10 Pack | 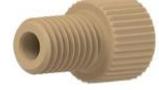 |

## SIF 3. UV characterisation

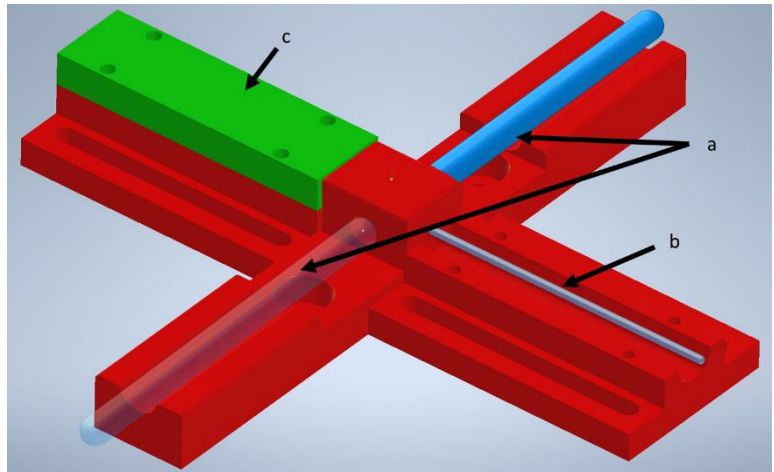
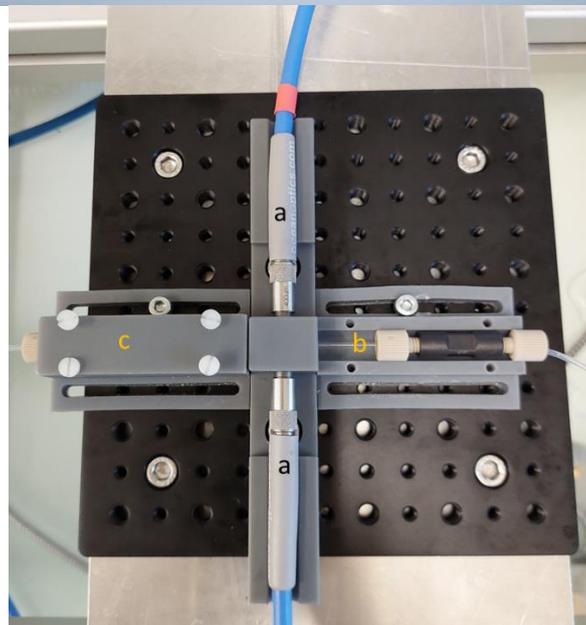

*SIF Figure 1 : UV holder: The central part serves to hold UV-solarised optical fiber (a) aligned on both sizes of the quartz capillary (b). A printed cover (c) is added on both sides to prevent the capillary moving and breaking.*

## SIF 4. Raman probe holder

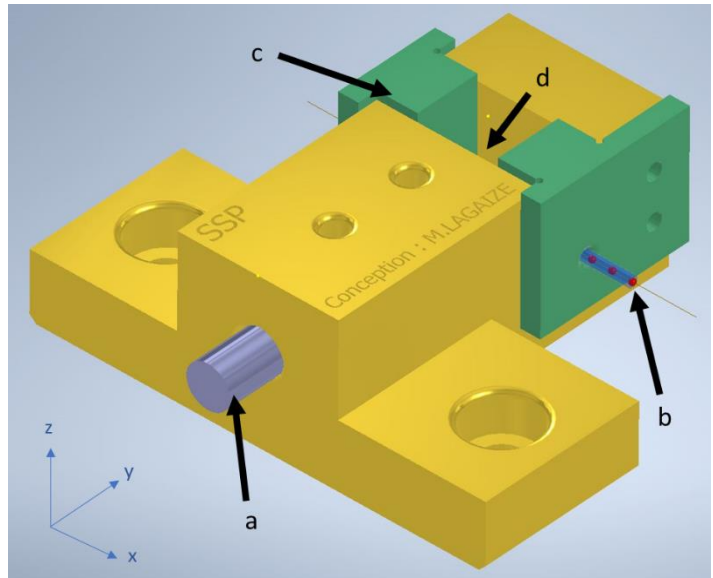

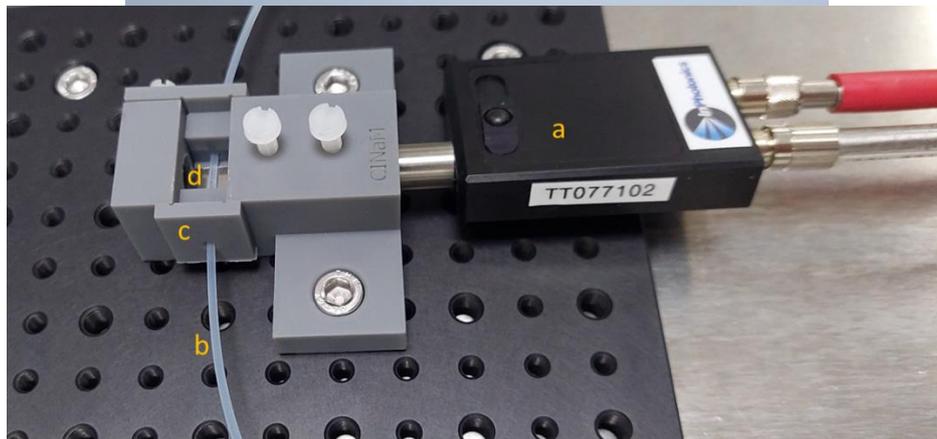

*SIF Figure 2: Raman probe holder: the Raman probe (a) position can be adjusted in y, to focus the beam on the middle of the capillary (b). The capillary is hold by a mobile piece (c) to adjust its position in z. It can also manually be moved in x. A mirror is fixed in the bottom of the observation zone (d).*

## SIF 5. Droplet UV analysis

In the case of saturated droplets, the UV cell is heated by an infra-red lamp to prevent droplet crystallisation. In this case, the absolute value of intensity is not considered for droplet size and frequency determination. The signal is analysed by peak detection on Python. As the collected data represent droplets which are deformed by their movement, their size is slightly overestimated. However, the measurements can still be used to determine the droplet size dispersion and distinguish quasi-spherical droplets from plugs.

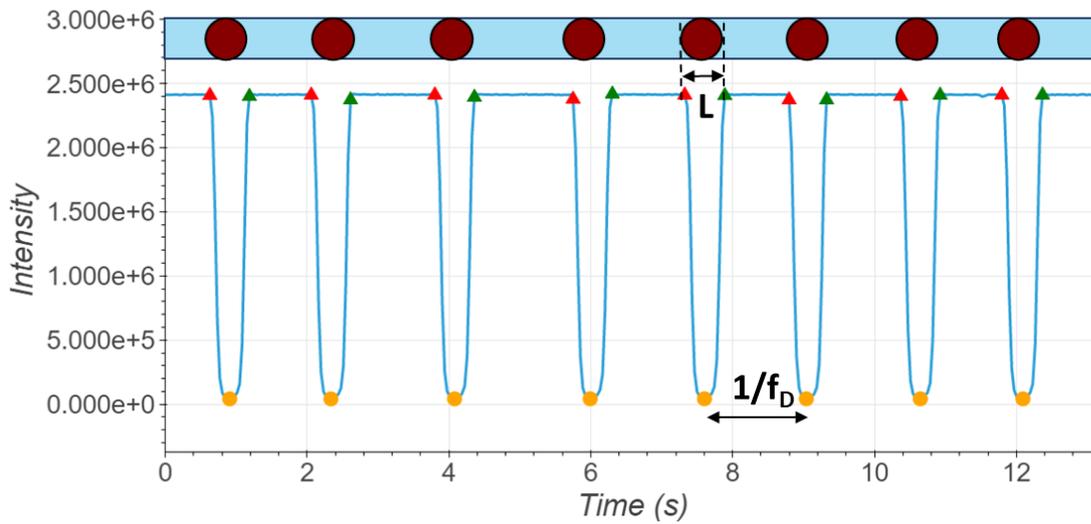

SIF Figure 3 : Real-time recording of light intensity at 280 nm of circulating droplets. L and $f_D$ are respectively the length of the droplets and their frequency.

## SIF 6. Raman spectra of Rimonabant in acetonitrile droplets (with and without crystal), GPL107 oil and references on powder (Forms I and II)

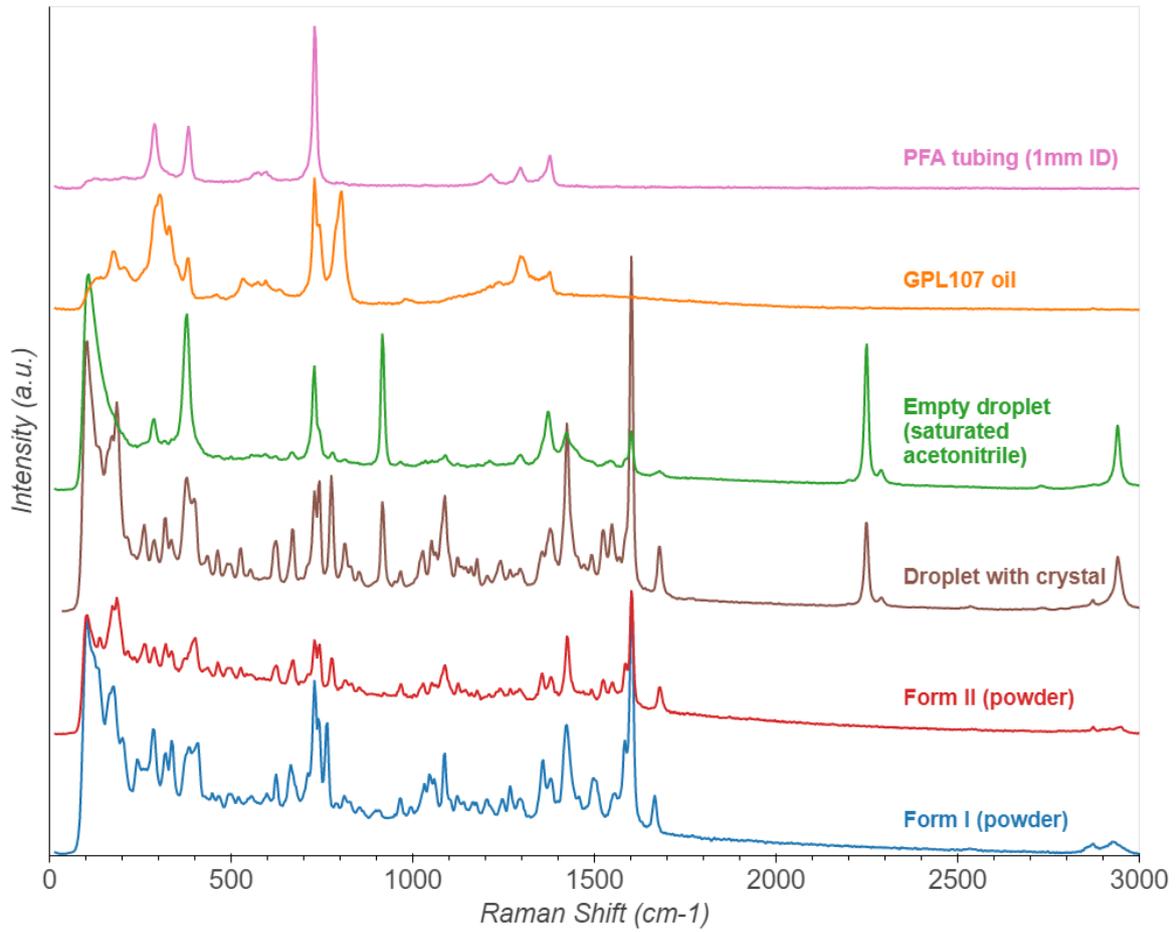

*SIF Figure 4 : Raman spectra of GPL 107 oil, an empty droplet of Rimonabant in acetonitrile, a droplet of acetonitrile containing a Rimonabant crystal (form II) and references of Rimonabant form I and II from powder*

**SIF 7. Solubility curves of Irbesartan, Rimonabant and Aripiprazole measured with microfluidics**

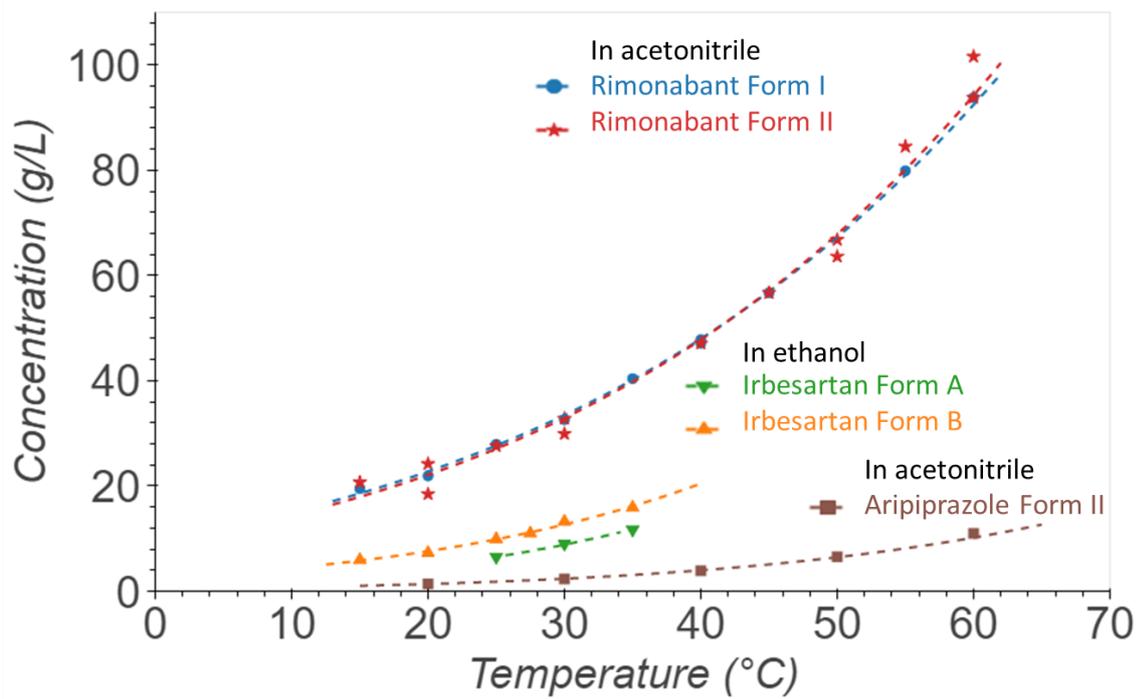

*SIF Figure 5: Solubility curves of Irbesartan in ethanol, Rimonabant in acetonitrile and Aripiprazole in acetonitrile measured with microfluidics.*

**SIF 8.    UV signal as a function of time of the diluted solution of Aripiprazole in acetonitrile.**

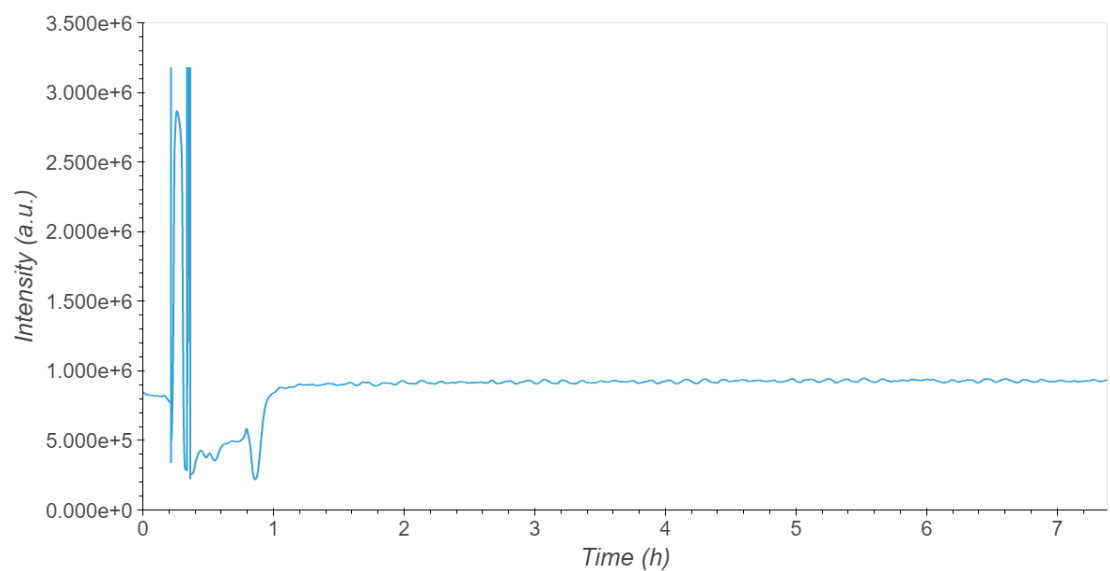

*SIF Figure 6 : UV signal as a function of time of the diluted solution of Aripiprazole in acetonitrile*

## SIF 9. Droplets optical characterisation

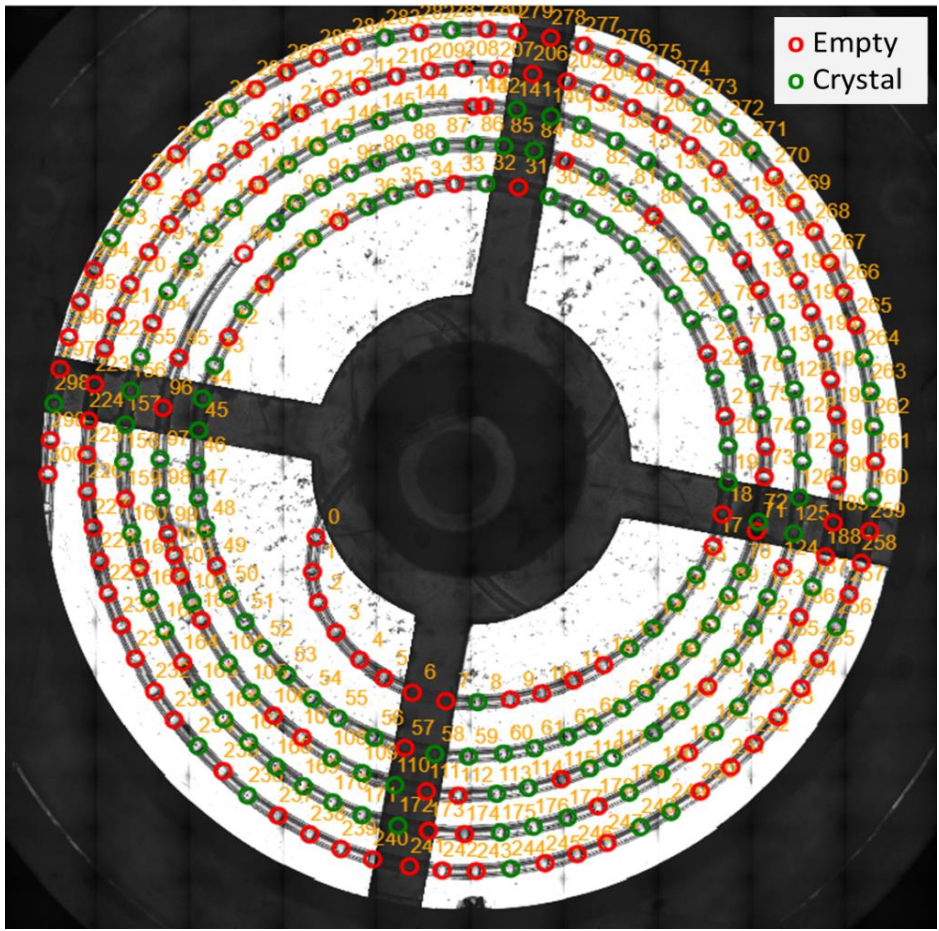

*SIF Figure 7 : Reconstructed picture of the spiral for the last picture cycle for Sulfathiazole in water (ramp -5°C/h) after 3.5 days (3 days at 10°C).*

## SIF 10. XRD patterns of Sulfathiazole

XRD measurements were realized in transmission mode using a high brilliancy rotating anode, Rigaku RU-200BH, (operating power 50 kV - 50 mA) equipped with a double reflexion mirror, Osmic, and an image plate detector, Mar345. The radiation used is the Cu K$\alpha$, $\lambda$ = 1.5418Å, and the beam size 0.5 x 0.5 mm$^2$. The maximum measurable 2$\theta$ angle is 65°, this is limited by the size of the detector and the minimum distance between the sample and the detector, the experimental resolution is about 0.3° in 2$\theta$. For the measurements the crystals were collected from the droplets and introduced into Lindeman glass capillaries, typically 0.5 mm diameter.

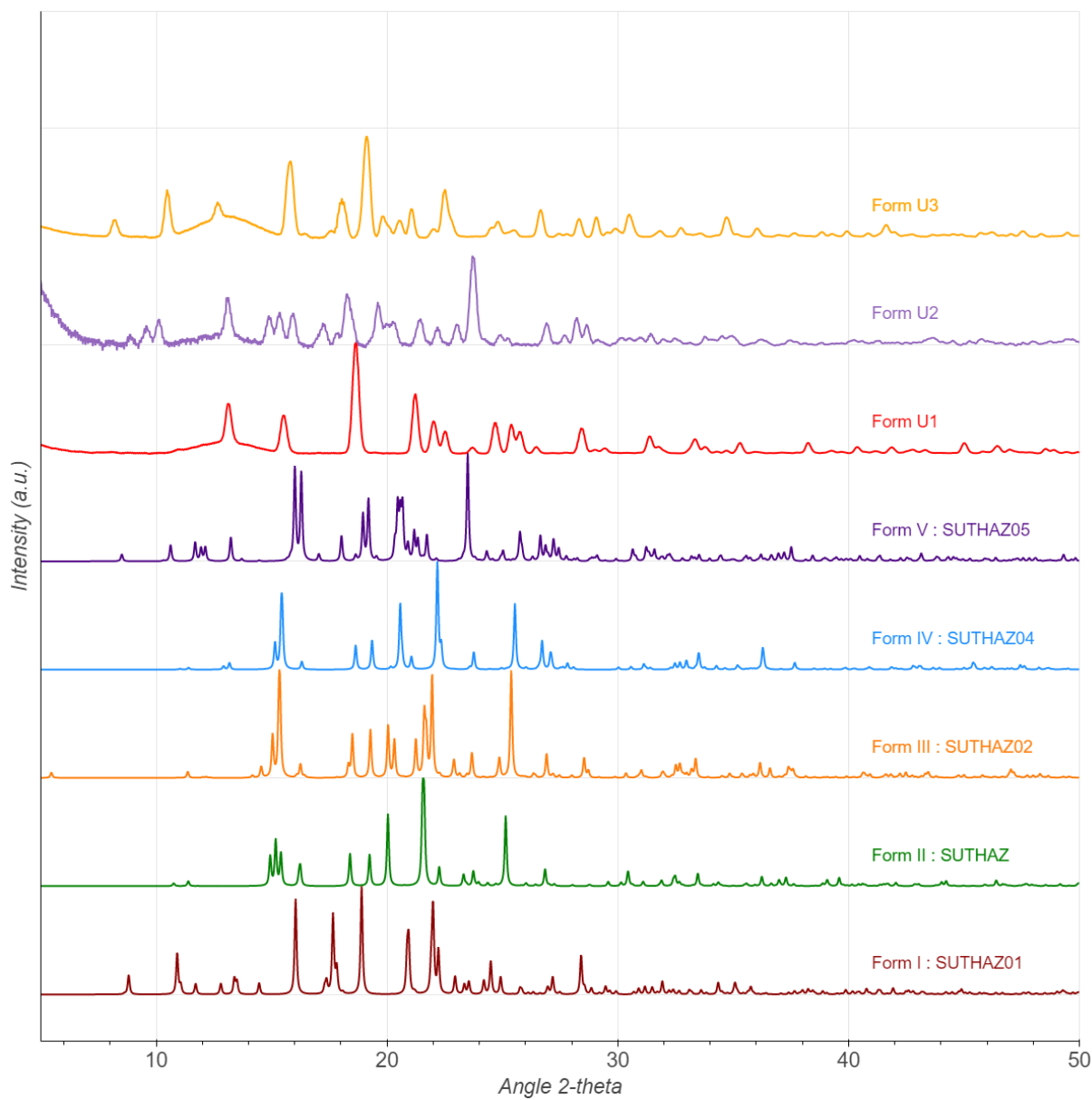

*SIF Figure 8 : XRD patterns of different crystalline forms of Sulfathiazole. SUTHAZ, SUTHAZ01, SUTHAZ02, SUTHAZ04 and SUTHAZ05 are references from the Cambridge Structural Database. U1, U2 and U3 were measured on crystals obtained in water droplets.*

## SIF 11. Raman spectra of Sulfathiazole in water droplets

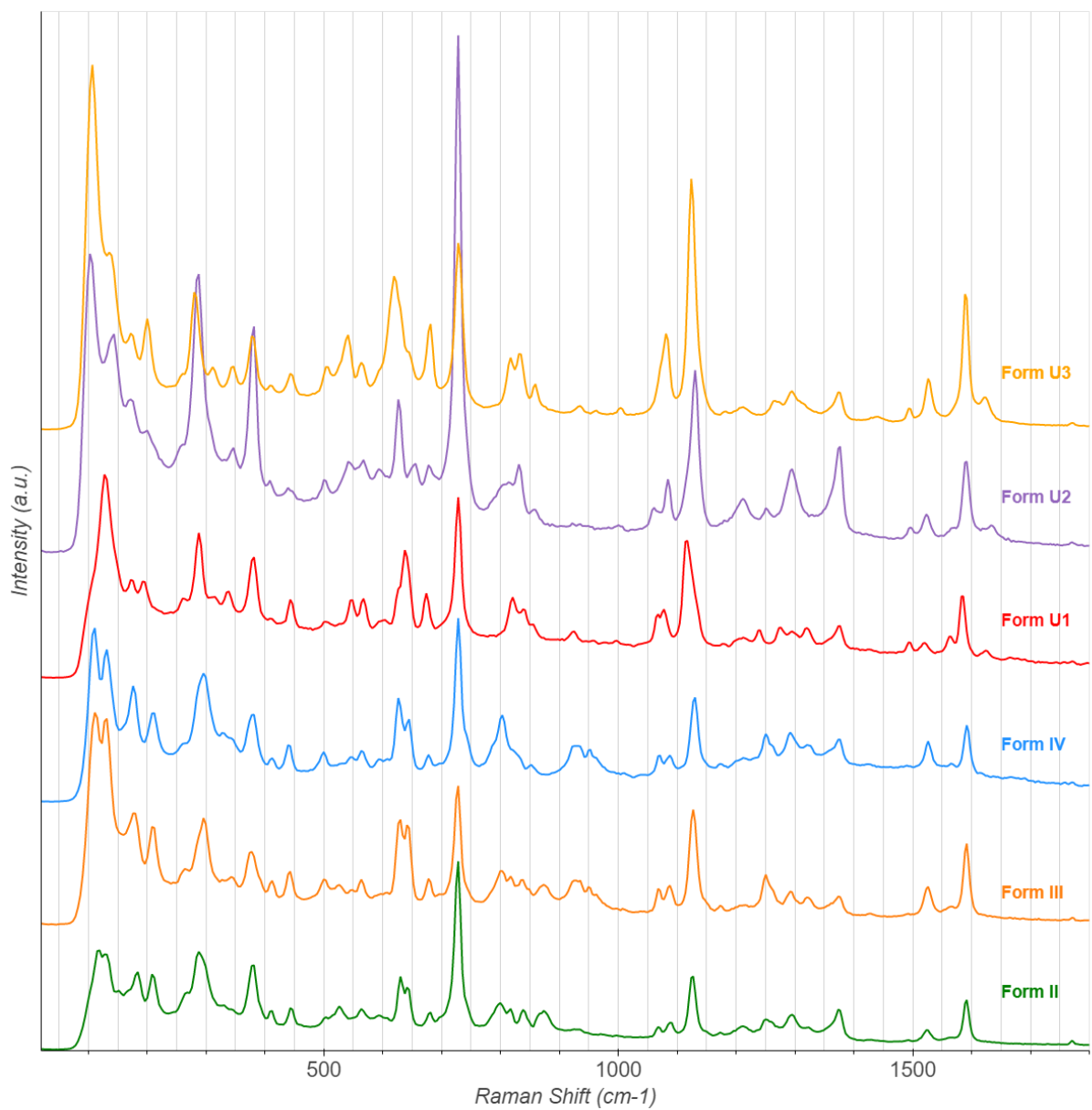

*SIF Figure 9: Raman spectra of different crystalline forms of Sulfathiazole, measured on crystals obtained in water droplets with QEPro Raman spectrometer, 785nm.*

## SIF 12.    Dissolution experiment

A slice of the tubing of the fast-cooling crystallisation of Sulfathiazole in water (-5°C/h) was used for this experiment. Two crystals in neighbouring droplets were analysed by Raman spectroscopy as form IV and U1 respectively. The capillary was placed in the thermostatic bath at 25°C, and progressively heated to 72°C. The two selected droplets were observed during the heating process. Pictures were taken every 2 to 4 minutes to monitor crystal dissolution. The U1 crystal completely dissolved around 65°C, whereas the crystal of form IV dissolved around 70°C.

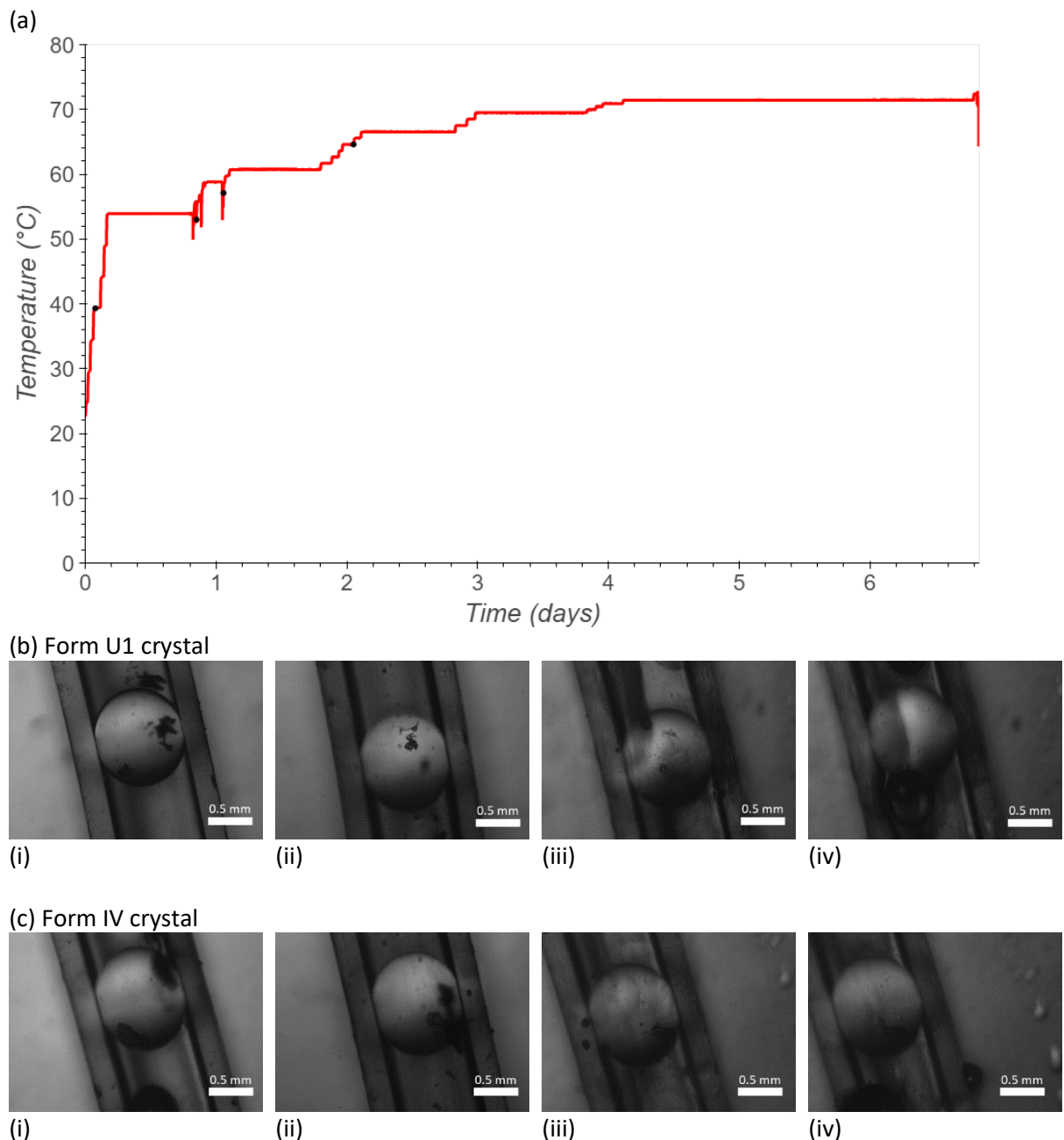

*SIF Figure 10: (a) Temperature profile during the dissolution experiment of Sulfathiazole forms U1 and IV in water. The black dots indicate the pictures represented in (b) and (c) for both crystals. (b) and (c) Pictures taken during the dissolution of crystal form U1 (b) and form IV (c) in water at different temperatures: (i) 39.3°C, (ii) 53.0°C, (iii) 57.1°C, (iv) 65.6°C. Scale bar is 0.5mm.*

**SIF 13.      Crystal habits of Sulfathiazole in acetonitrile**

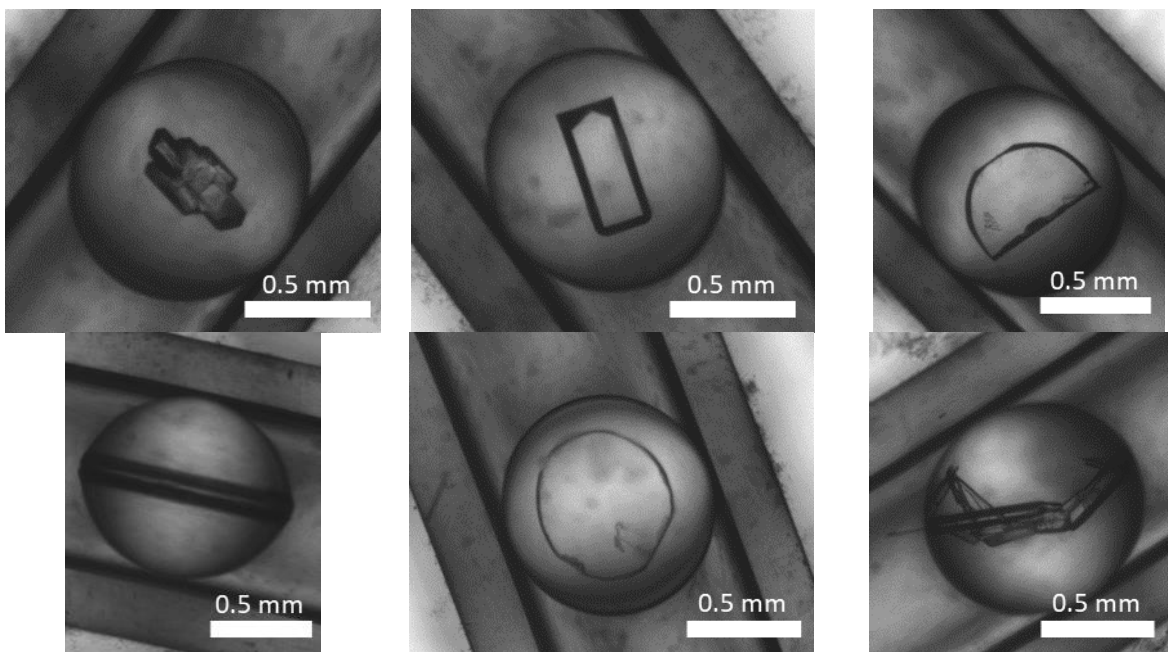

*SIF Figure 11: Crystals habits of Sulfathiazole obtained in acetonitrile with a -5°C ramp from 70°C to 10°C*

35